\documentclass[prd,showpacs,preprintnumbers,amsmath,amssymb]{revtex4}

\usepackage{graphicx}
\usepackage{dcolumn}
\usepackage{bm}

\begin{document}

\title{Conformally Coupled Induced Gravity as an Infrared Fixed Point}

\author{Yongsung Yoon}\email{cem@hanyang.ac.kr}
\affiliation{Physics Department, Research Institute for Natural Sciences, Hanyang University, Seoul 133-791, Korea}

\date{\today}

\begin{abstract}
We have found that the conformally coupled induced gravity can be an infrared fixed point of induced gravity with
Yukawa couplings with matter. The late time cosmology with a uniform mean matter distribution can be described by the
conformally coupled induced gravity, which has an emergent global conformal symmetry in the cosmic scale. Aiming to
resolve the puzzles for the dark energy, we have obtained exact cosmological equations and determined the dark energy
density, the matter density, and the jerk parameter in the present universe based on the recent observational cosmic
expansion data for $a/H^{2}$.
\end{abstract}
 \maketitle

General Relativity provides a set of remarkably successful Friedmann equations to our modern cosmology. Since the
discovery of Hubble expansion of our universe, for many decades it was believed that the cosmological constant was zero
mysteriously, even though a uniform vacuum energy could be generated by quantum fields. However, the recent
astronomical observations with the help of Type-I supernovae tell us that the cosmological constant is not zero, but so
small that the dark energy density is quite close to the matter density at present \cite{ACC1,ACC2,ACC3}. Thus, we are
now confronting two puzzles for the dark energy. The old puzzle is how the vacuum energy, which could be generated by
quantum fields, could not contribute to the dark energy. The new puzzle is why the dark energy density is so small and
close to the matter density at present coincidently. However, in General Relativity, there seems to be no room to
conceive any mechanism which can resolve these puzzles.

Before the success of Weinberg-Salam model, the weak interactions were characterized by the dimensionful Fermi's
coupling constant $G_{F}$. But later, the Fermi's coupling constant is replaced with the Higgs scalar field in Standard
Model. Finally, the Higgs particle seems to have been found recently \cite{Higgs}. From the lesson of this success, it
is quite tempting to replace the dimensionful Newton's constant $G_{N}$ with a new gravity scalar field as Brans-Dicke
gravity model \cite{BD1,BD2}, or induced gravity models \cite{IGO1,IGO2,IGO3,IGO4,IG1,IG2,IG3,IG4,IG5,IG6,IG7}.

Thus, replacing the Newton's constant $G_{N}$ with a gravity scalar field $\Phi$ as $8\pi G_{N}=1/\Phi^{2}$, we
consider the following induced gravity action with a gravity scalar potential $V(\Phi)$,
\begin{eqnarray}
S= -\int d^{4}x\sqrt{-g}\left[\frac{\xi}{2}\Phi^{2} R +\frac{1}{2}(\partial
\Phi)^{2}\right] + \int d^{4}x\sqrt{-g} \left({\cal L}_{m} - V(\Phi) \right),
\label{action}
\end{eqnarray}
where ${\cal L}_{m}$ is the matter Lagrangian, and $\xi$ is a dimensionless coupling constant between the gravity
scalar field and the metric. $\xi=1/6$ is for the conformal coupling \cite{JYP}.

The equations of motion for the gravity scalar field and the metric can be written as
\begin{eqnarray}
\nabla^{2}\Phi = \xi\Phi R + \frac{\partial V(\Phi)}{\partial \Phi} - J_{\Phi}, \label{S0-eq}
\end{eqnarray}
\begin{eqnarray}
\xi\Phi^{2} G_{\mu\nu} - 2\xi\Phi\nabla_{\mu}\nabla_{\nu}\Phi +2\xi g_{\mu\nu}\Phi\nabla^{2}\Phi
+(1-2\xi)\nabla_{\mu}\Phi \nabla_{\nu}\Phi - \frac{1-4\xi}{2}g_{\mu\nu}(\nabla\Phi)^{2} - g_{\mu\nu}V(\Phi) =
T^{(m)}_{\mu\nu} , \label{G0-eq}
\end{eqnarray}
where $T^{(m)}_{\mu\nu}$ is the matter energy-momentum tensor, and the conformal charge density is define as $J_{\Phi}
\equiv \partial{\cal L}_{m}/\partial\Phi$. Combining the trace part of equation (\ref{G0-eq}) with the gravity scalar
equation (\ref{S0-eq}), we can rewrite the gravity scalar equation equivalently as
\begin{eqnarray}
(6\xi-1)[\Phi \nabla^{2}\Phi + (\nabla\Phi)^{2}]  = T^{(m)} +4V(\Phi) -\Phi\frac{\partial V(\Phi)}{\partial \Phi} + J
_{\Phi} , \label{DGS-eq}
\end{eqnarray}
where $T^{(m)}$ is the trace of the matter energy-momentum tensor.

As we can see from Eq.(\ref{DGS-eq}), the driving force of the gravity scalar field is given by the conformal anomaly,
$T^{(m)} +4V -\Phi \partial V(\Phi)/\partial\Phi +J_{\Phi}$, of the whole system, which is quite distinctive from an
ordinary scalar field whose driving force is the gradient of its potential. Thus, the gravity scalar field can not be
settled down at a potential minimum, $\partial V(\Phi)/\partial\Phi=0$, unlike an ordinary scalar field, but be driven
to $T^{(m)} +4V -\Phi \partial V(\Phi)/\partial\Phi+J_{\Phi}=0$ with the emergent conformal symmetry.

Due to the presence of the gravity scalar potential $V(\Phi)$, the equations Eqs.(\ref{S0-eq},\ref{G0-eq}) might not
allow the exact flat spacetime solution for a constant gravity scalar field even in vacuum. However, in a non-cosmic
scale, the deviation from the flat spacetime could be negligible provided that the gravity scalar potential satisfies a
certain condition.

We may assume a Yukawa coupling between the gravity scalar field and a spinor matter field, assigning an internal
charge to the gravity scalar field if necessary, like $i h f_{abc}\bar{\psi}^{a}\Phi^{b}\psi^{c}$. Then, one-loop
renormalization gives the renormalization group equation for the coupling $\xi$ as
\begin{eqnarray}
\mu\frac{d\xi}{d\mu} = (\xi-\frac{1}{6}) C \ h^{2} \label{RG-eq},
\end{eqnarray}
where $h$ is the Yukawa coupling constant, and C is a positive constant which depends on the internal group
\cite{buch}.

We are interested in the infrared limit, not in the ultra-violet limit. If $\xi
> 1/6$ at high energy, as the energy
scale $\mu$ gets lower, the RG-equation drives $\xi$ smaller to the infrared fixed point $\xi=1/6$. If $\xi < 1/6$ at
high energy, the RG-equation drives $\xi$ larger, as the energy scale $\mu$ gets lower, to the infrared fixed point
$\xi=1/6$ also at low energy. Thus, the conformal coupling $\xi=1/6$ is the infrared fixed point of the induced gravity
for any $\xi$. Thus, when we examine the late time cosmology with a low matter density, which is a far infrared limit,
it is appropriate to deal with the conformal coupling, $\xi=1/6$. In the cosmic scale, integrating out all locally
fluctuating matter fields, we have the conformally coupled induced gravity with a mean uniform matter energy momentum
tensor. However, for the gravity in the early universe at high energy or in a non-cosmic scale with locally fluctuating
matter fields, we could not assume the conformal coupling $\xi=1/6$ in induced gravity.

To examine the gravity in the cosmic scale with a uniform matter distribution, we take the conformal coupling
$\xi=1/6$. Then the conformally coupled induced gravity action can be rewritten as
\begin{eqnarray}
S= -\int d^{4}x\sqrt{-g}\left[\frac{1}{2}\Phi^{2} R +3(\partial
\Phi)^{2}\right] + \int d^{4}x\sqrt{-g} \left({\cal L}_{m} - V(\Phi) \right)
\label{action} .
\end{eqnarray}
The first gravity part of the action (\ref{action}) is invariant under the
local conformal transformation, $g_{\mu\nu}(x) \rightarrow e^{2\sigma(x)}
g_{\mu\nu}(x), \quad \Phi(x) \rightarrow e^{-\sigma(x)} \Phi(x)$, which is the
reason why we call the action (\ref{action}) as the conformally coupled
gravity. However, the second matter part of the action (\ref{action}) does not
have any explicit conformal symmetry at all as we know.

We have the equations of motion for the gravity scalar field $\Phi$ and the metric $g_{\mu\nu}$ as follows,
\begin{eqnarray}
6\nabla^{2}\Phi = \Phi R + \frac{\partial V(\Phi)}{\partial \Phi} , \label{S-eq}
\end{eqnarray}
\begin{eqnarray}
\Phi^{2} G_{\mu\nu} - 2\Phi\nabla_{\mu}\nabla_{\nu}\Phi
+2g_{\mu\nu}\Phi\nabla^{2}\Phi +4\nabla_{\mu}\Phi \nabla_{\nu}\Phi -
g_{\mu\nu}(\nabla\Phi)^{2} - g_{\mu\nu}V(\Phi) = T^{(m)}_{\mu\nu} .
\label{G-eq}
\end{eqnarray}

Taking the trace of the metric equation (\ref{G-eq}) gives an expression of
$\Phi^{2}R$. Plugging it into Eq.(\ref{S-eq}), we have a characteristic
equation of the conformally coupled induced gravity,
\begin{eqnarray}
T^{(m)} +4V(\Phi) -\Phi\frac{\partial V(\Phi)}{\partial \Phi} = 0 , \label{GS-eq}
\end{eqnarray}
where $T^{(m)}$ is the trace of a uniform matter energy-momentum tensor. The origin of the constraint (\ref{GS-eq}) is
that the gravity scalar field $\Phi$ is not dynamical, but auxiliary in the conformally coupled induced gravity
\cite{JYP}. The gravity scalar field settles down not at a potential minimum, $\partial V(\Phi)/\partial\Phi =0$, but
at the point of the emergent conformal symmetry satisfying the equation (\ref{GS-eq}).

Even though the classical matter action, the second part of (\ref{action}), does not have an explicit conformal
symmetry in general, the equations of motion (\ref{S-eq},\ref{G-eq}) requires the global conformal symmetry in the
whole system. To see this more clearly, let us consider the following global conformal transformation with a constant
real parameter $\lambda$,
\begin{eqnarray}
g_{\mu\nu}(x) \rightarrow e^{2\lambda} g_{\mu\nu}(x), \quad \Phi(x) \rightarrow
e^{-\lambda} \Phi(x) \label{scale-eq} .
\end{eqnarray}

If we assume the invariance of the action (\ref{action}) under this global conformal transformation (\ref{scale-eq}),
we have the following equation for the invariance,
\begin{eqnarray}
0 = \delta(\sqrt{-g}{\cal L}_{m}-\sqrt{-g}V(\Phi)) = -\sqrt{-g}\left( T^{(m)}
+4V(\Phi ) - \Phi\frac{\partial V(\Phi)}{\partial\Phi} \right)\delta\lambda
\label{trf-eq} ,
\end{eqnarray}
where we have used $\delta(\sqrt{-g}{\cal
L}_{m})=-\sqrt{-g}T^{(m)}\delta\lambda$ with an infinitesimal parameter
$\delta\lambda$. The equation (\ref{trf-eq}) is exactly the trace anomaly
relation (\ref{GS-eq}).

Therefore we have found that, without any explicit conformal symmetry in the classical matter Lagrangian, the
conformally coupled induced gravity, which is the infrared fixed point of induced gravity, exhibits an emergent global
conformal symmetry (\ref{scale-eq}) in the whole system of the cosmic scale. When we write down an explicit form of the
matter energy momentum tensor, the metric, or the gravity scalar field, it means that a scale has been chosen with a
fixed conformal gauge parameter $\lambda$.

Because the conformally coupled gravity is appropriate to describe the gravity in the cosmic scale with a uniform
matter distribution, we investigate the cosmological evolution, assuming a homogeneous gravity scalar field $\Phi(t)$
depending only on time, and adopting the Robertson-Walker metric with a vanishing spatial curvature($k=0$),
\begin{eqnarray}
ds^{2}=dt^{2}-S^{2}(t)[dr^{2}+r^{2}d\Omega^{2}]. \label{metric}
\end{eqnarray}

Then, Eqs.(\ref{G-eq},\ref{GS-eq}) are reduced to a set of cosmological
equations as shown below. Denoting the time derivative as an over-dot,
\begin{eqnarray}
3\Phi^{2}H^{2} +6H\Phi\dot{\Phi} +3\dot{\Phi}^{2} = \rho_{m} +
\rho_{\Lambda}(\Phi) , \label{CosG00-eq}
\end{eqnarray}
\begin{eqnarray}
2 a \Phi^{2} +H^{2}\Phi^{2} +4H\Phi\dot{\Phi} +2\Phi\ddot{\Phi} -
\dot{\Phi}^{2} = -p_{m} + \rho_{\Lambda}(\Phi) , \label{CosG11-eq}
\end{eqnarray}
\begin{eqnarray}
\rho_{m} - 3p_{m} +4\rho_{vac} +4V(\Phi) = \Phi\frac{\partial V(\Phi)}{\partial
\Phi} , \label{CosD-eq}
\end{eqnarray}
where $H \equiv \dot{S}/S$ is the Hubble parameter and $a \equiv \ddot{S}/S$ is the cosmic acceleration parameter. With
the matter pressure $p_{m}$ and the matter density $\rho_{m}$, $~T^{(m)0}_{0}=\rho_{m}+\rho_{vac}$ and
$~T^{(m)1}_{1}=T^{(m)2}_{2}=T^{(m)3}_{3}=-p_{m}+\rho_{vac}$, where $\rho_{vac}$ is the vacuum energy which could be
generated by quantum effects in the matter Lagrangian. We can see that the sum of the vacuum energy $\rho_{vac}$ in the
matter Lagrangian and the cosmic potential $V(\Phi)$ plays the role of the total dark energy density,
$\rho_{\Lambda}(\Phi) \equiv \rho_{vac} + V(\Phi)$.

We try to find a cosmic scalar potential in the conformally coupled gravity which can relax the uniform vacuum energy
which could be generated in the matter Lagrangian. For a given vacuum energy $\rho_{vac}$ with the matter density
$\rho_{m}$ and the matter pressure $p_{m}$ from the matter Lagrangian, a gravity scalar field $\Phi$ would satisfy
Eq.(\ref{CosD-eq}). For a different vacuum energy $\rho'_{vac}$, the equation (\ref{CosD-eq}) would be satisfied with a
different gravity scalar field $\Phi'$ instead of $\Phi$. For these two different vacuum energies $\rho_{vac}$ and
$\rho'_{vac}$, the difference in the total dark energy densities is found as
\begin{eqnarray}
\Delta\rho_{\Lambda} =\rho_{\Lambda}(\Phi') -\rho_{\Lambda}(\Phi) =
V(\Phi')-V(\Phi)+\rho'_{vac}-\rho_{vac} = \frac{1}{4} \left[
\Phi'\frac{\partial V(\Phi)}{\partial \Phi}\mid_{\Phi'} - \Phi\frac{\partial
V(\Phi)}{\partial \Phi}\mid_{\Phi} \right] . \label{Zero-eq}
\end{eqnarray}

Therefore, we can have the same total dark energy density $\rho_{\Lambda}$ in the conformally coupled gravity whatever
a uniform vacuum energy $\rho_{vac}$ generated in matter Lagrangian is, provided that the cosmic scalar potential
satisfies the equation, $\Phi\partial V(\Phi)/\partial\Phi = \alpha$ for a constant $\alpha$. Thus we can determine the
cosmic scalar potential uniquely as;
\begin{eqnarray}
V(\Phi) = \frac{\alpha}{2}\ln(\frac{\Phi}{\Phi_{0}})^{2} , \quad \rho_{\Lambda}(\Phi) = \rho_{vac} +
\frac{\alpha}{2}\ln(\frac{\Phi}{\Phi_{0}})^{2} \label{Pot-eq} ,
\end{eqnarray}
where $\alpha$ and $\Phi_{0}$ are dimensionful constants.

For the cosmic potential Eq.(\ref{Pot-eq}), the total dark energy density
$\rho_{\Lambda}$ satisfies the constraint equation (\ref{CosD-eq}) with the
matter density $\rho_{m}$ and the matter pressure $p_{m}$,
\begin{eqnarray}
\rho_{m} - 3p_{m} + 4\rho_{\Lambda}(\Phi) = \alpha . \label{CR-eq}
\end{eqnarray}
The left hand side of this equation is the trace of the total energy momentum tensor including all matter and total
dark energy of our universe. If we interpret Eq.(\ref{CR-eq}) as the trace anomaly cancellation equation which states
that the total trace anomaly of our universe vanishes, then the constant $-\alpha$ in Eq.(\ref{CR-eq}) would be the
conformal anomaly density which is generated in matter Lagrangian by quantum effects. Thus, we may speculate that the
cosmic potential (\ref{Pot-eq}) would be an effective potential to describe the conformal anomaly generated in matter
Lagrangian by quantum effects. It is found that the cosmic potential (\ref{Pot-eq}) is the very linear form of coupling
$\phi\Theta^{\mu}_{\mu}$ between a scalar field $\phi$ and a conformal anomaly $\Theta^{\mu}_{\mu}$, which might have
an origin in quantum chromodynamics, considered in \cite{anomal} with the field redefinition, $\Phi \equiv
e^{-\phi}/\kappa_{0}$. Thus, we may call $\alpha$ a conformal anomaly parameter.

The gravitational constant $G_{N,vac}=1/(8\pi\Phi^{2}_{vac})$ in vacuum is determined from
Eqs.(\ref{Pot-eq},\ref{CR-eq}) as
\begin{eqnarray}
\rho_{\Lambda,vac} \equiv \rho_{vac} + \frac{\alpha}{2}\ln(\frac{\Phi_{vac}}{\Phi_{0}})^{2}=\frac{1}{4}\alpha
 \label{vac-eq} .
\end{eqnarray}
Both the dimensionful constant $\Phi_{0}$ and the vacuum energy $\rho_{vac}$ hide into the dark energy density
$\rho_{\Lambda}(\Phi)$ as Eq.(\ref{Pot-eq}), giving the gravitational constant $G_{N,vac}$ in vacuum, and never appear
as physical quantities at all.

Taking a time derivative Eq.(\ref{CosG00-eq}), and using
Eqs.(\ref{CosG11-eq},\ref{CosD-eq}), we have the exact matter energy
conservation equation,
\begin{eqnarray}
\frac{d}{dt}\rho_{m} + 3H(\rho_{m}+p_{m}) = 0 , \label{matter-eq}
\end{eqnarray}
which is independent of the gravity scalar field $\Phi$. Therefore, for the matter with the equation of state $w$ such
that $p_{m}=w\rho_{m}$, the evolution of matter density,
\begin{eqnarray}
\rho_{m}  = \frac{A}{S^{3(w+1)}(t)} , \label{rho-eq}
\end{eqnarray}
is not affected at all by the gravity scalar in the induced gravity. The gravity scalar field evolves slowly due to the
change of matter density

\begin{eqnarray}
\frac{d\Phi}{dt} = \frac{3(1-3w)(1+w)}{4} H \Phi \frac{\rho_{m}}{\alpha} \label{mphi-eq},
\end{eqnarray}
as we can see from Eqs.(\ref{CR-eq},\ref{matter-eq}).

However, in the radiation dominated early universe, $\Phi$ takes the constant vacuum value $\Phi_{vac}$, as we can see
from Eqs.(\ref{CR-eq},\ref{vac-eq}), observing $p_{m,rad}=\rho_{m,rad}/3$ for relativistic matter. Therefore, in the
radiation dominated early universe, the cosmological evolution equations of the conformally coupled induced gravity is
reduced to the Friedmann equations with the constant gravitational constant $G_{N,vac}$ and the dark energy density
$\rho_{\Lambda,vac}$.

The dark energy density $\rho_{\Lambda}$ also evolves as the gravity scalar field evolves,
\begin{eqnarray}
\frac{d{\rho}_{\Lambda}}{dt} = \frac{3(1-3w)(1+w)}{4} H \rho_{m} \label{mdark-eq} .
\end{eqnarray}

Therefore, we can summarize the cosmological evolution of the conformally coupled induced gravity as follows. In the
radiation dominated early universe, the conformally coupled induced gravity gives the Friedmann equations of General
Relativity with the dark energy density $\alpha/4$, having a constant gravity scalar field $\Phi_{vac}$. In the late
time cosmological evolution of the matter dominated universe, it is believed that the equation of state for matter is
$w=0$ \cite{LCDM}. As the universe transits from the radiation dominated era with $w=1/3$ to the matter dominated era
with $w=0$, the trace of matter energy momentum tensor increases from 0 to $\rho_{m}$. The gravity scalar field
decreases to $\Phi_{vac}e^{-\rho_{m}/4\alpha}$, and the dark energy density also decreases to $\alpha-\rho_{m}$.
Finally, in the matter dominated era with $w=0$, while the matter density decreases as Eq.(\ref{rho-eq}), the gravity
scalar field $\Phi$ increases toward the vacuum value $\Phi_{vac}$ as Eq.(\ref{mphi-eq}), and the dark energy density
increases also toward the vacuum value $\alpha/4$ as Eq.(\ref{mdark-eq}).

If the matter density $\rho_{m}$ exceeds $\alpha$ in the early stage of the matter dominated era, the dark energy
density could be negative for a while.

Inserting Eq.(\ref{mphi-eq}) into the cosmological equations (\ref{CosG00-eq},\ref{CosG11-eq}), we have a set of exact
cosmological evolution equations in the matter dominated late time universe as follows,
\begin{eqnarray}
3\Phi^{2}H^{2} = \alpha\frac{1+3(1+w)x}{4(1+\nu x)^{2}} ,\label{H2-eq}
\end{eqnarray}
\begin{eqnarray}
6\Phi^{2}a = 3\alpha \frac{1-(1+w)x}{4(1+\nu x)} -\alpha\frac{(1+3(1+w)x)(1-2(2+3w)\nu x+\nu^{2}x^{2})}{4(1+\nu x)^{3}}
, \label{a-eq}
\end{eqnarray}
where we have defined $x \equiv \rho_{m}/\alpha$, $\nu \equiv \frac{3}{4}(1+w)(1-3w)$, and used
$\rho_{\Lambda}(\Phi)=(1-(1-3w)x)\alpha/4$ from Eq.(\ref{CR-eq}). The cosmological equations (\ref{H2-eq},\ref{a-eq})
approximate to the Friedmann equations only if $x$ is much smaller than $1$, i.e. $x \ll 1$.

The jerk parameter, a scaled third time derivative of the scale factor of the universe, is obtained as
\begin{eqnarray}
j \equiv \frac{1}{S(t)}\frac{d^{3}S(t)}{dt^3} = \frac{\dot{a}}{H^{3}} + \frac{a}{H^{2}} = \nonumber ~\\ 1 - \nu
x\frac{3(1+w)(4+3w)-(8+9w)\nu x+\nu^{2}x^{2}}{1+\nu x} + \frac{9}{2}(1+w)^{2}x\frac{1+\nu x}{1+3(1+w)x} +
\frac{a}{H^{2}}3\nu x\frac{2+3w-\nu x}{1+\nu x} \label{j-eq} ,
\end{eqnarray}
which depends only on $x$ and $w$ because $a/H^{2}$ is also a function of $x$ and $w$ only from
Eqs.(\ref{H2-eq},\ref{a-eq}).

Recently, the ratio of the cosmic acceleration parameter $a$ to the squared Hubble parameter $H^{2}$ has been
determined with the help of Type-I supernovae as standard candles \cite{ACC1,ACC2,ACC3},
\begin{eqnarray}
\left(\frac{a}{H^{2}}\right)_{exp} \cong 0.55 \label{ratio-exp}.
\end{eqnarray}

From the ratio of Eq.(\ref{a-eq}) to Eq.(\ref{H2-eq}), $a/H^{2}$, which is a monotonically decreasing function of $x$,
we can find a solution $x$ satisfying the experimental ratio $(a/H^{2})_{exp}$ numerically. If we use the experimental
value (\ref{ratio-exp}), then we find the corresponding solution $x=0.476$ for $w=0$. For $x=0.476$, the cosmological
equations (\ref{H2-eq},\ref{a-eq}) do not approximate to the Friedmann equations because the higher $x$ terms are not
negligible. This $x=0.476$ determines the matter density $\rho_{m}$, the dark energy density $\rho_{\Lambda}$, the jerk
parameter $j$, and the critical density $\rho_{c}$, with $w=0$ at present, respectively as
\begin{eqnarray}
\rho_{m}=0.476\alpha, \quad \rho_{\Lambda}=0.13\alpha, \quad j= 0.47, \quad \rho_{c} \equiv 3\Phi^{2}H^{2} =
\frac{3}{8\pi G_{N}}H^{2} =0.33\alpha \label{fit-exp1}.
\end{eqnarray}
Thus, the ratio of the dark energy density to the matter density in the present universe is $\rho_{\Lambda}/\rho_{m}
\simeq 1/4$, which is quite smaller than the value $\rho_{\Lambda}/\rho_{m} \simeq 7/3$ determined by the Friedmann
equations based on the same data (\ref{ratio-exp}). The jerk parameter $j=0.47$ is in the current observational range,
$0 \lesssim j \lesssim 2$ \cite{jerk1,jerk2}. From the difference between the vacuum and the current dark energy
densities, it is found that $G_{N,now} = 1.27 G_{N,vac}$. The current gravitational constant is decreasing to the
vacuum value, restoring to the gravitational constant in the radiation dominated early universe.

If the equation of state for matter in our current universe is not $w=0$, but $w=1/10$ for example, then we have the
followings instead,
\begin{eqnarray}
\rho_{m}=0.221\alpha, \quad \rho_{\Lambda}=0.21\alpha, \quad j= 0.71, \quad \rho_{c} =0.34\alpha, \quad {\rm for~}
w=1/10 \label{fit-exp2}.
\end{eqnarray}

As conclusions, observing that the conformally induced gravity can be an infrared fixed point of induced gravity with
Yukawa couplings with matter, we have found that induced gravity with a locally fluctuating matter energy momentum
tensor in a non-cosmic scale yields the conformally coupled induced gravity with a uniform mean matter distribution at
the infrared limit in the cosmic scale. Using the conformally coupled gravity with the uniquely determined cosmic
potential, we have found that the dark energy density is determined to the order of the matter density. In the matter
dominated late time universe, we have a set of exact cosmological equations which deviate from the Friedmann equations
significantly. However, in the radiation dominated early universe, we regain the Friedmann equations of General
Relativity having a constant gravity scalar field. Based on the recent observational cosmic expansion data for
$a/H^{2}$, assuming the equation of state for matter $w=0$, we have determined the ratio of the dark energy density to
the matter density as about 1/4 in the present universe. This is quite smaller than the value $7/3$ determined by the
Friedmann equations.



\end{document}